\documentclass[runningheads]{llncs}
\usepackage{hyperref}
\usepackage{graphicx}
\usepackage{float}
\usepackage{todonotes}
\usepackage{ctable}

\begin{document}
\title{Editable Stain Transformation Of Histological Images Using Unpaired GANs}
\author{Tibor Sloboda\orcidID{0000-0001-6817-6297}\and\\
Lukáš Hudec\orcidID{0000-0002-1659-0362}\and\\
Wanda Benešová\orcidID{0000-0001-6929-9694}}

\institute{Vision \& Graphics Group,\\
Faculty of Informatics and Information Technology,\\
Slovak University of Technology,\\
Bratislava, Slovak Republic\\
\email{xslobodat2@stuba.sk}\\
\url{https://vgg.fiit.stuba.sk/}}

\maketitle              
\begin{abstract}
Double staining in histopathology is done to help identify tissue features and cell types differentiated between two tissue samples using two different dyes. In the case of metaplastic breast cancer, H\&E and P63 are often used in conjunction for diagnosis. However, P63 tends to damage the tissue and is prohibitively expensive, motivating the development of virtual staining methods, or methods of using artificial intelligence in computer vision for diagnostic strain transformation. In this work, we present results of the new xAI-CycleGAN architecture's capability to transform from H\&E pathology stain to the P63 pathology stain on samples of breast tissue with presence of metaplastic cancer. The architecture is based on Mask CycleGAN and explainability-enhanced training, and further enhanced by structure-preserving features, and the ability to edit the output to further bring generated samples to ground truth images. We demonstrate its ability to preserve structure well and produce superior quality images, and demonstrate the ability to use output editing to approach real images, and opening the doors for further tuning frameworks to perfect the model using the editing approach. Additionally, we present the results of a survey conducted with histopathologists, evaluating the realism of the generated images through a pairwise comparison task, where we demonstrate the approach produced high quality images that sometimes are indistinguishable from ground truth, and overall our model outputs get a high realism rating.
\end{abstract}
\newpage
\section{Introduction}
Histopathology images are often difficult to register properly, leading to issues in using paired approaches, which may yield better results. Often, there is also significant tissue damage caused by certain immunohistochemical stains used in particular to help pathologists identify the presence of cancerous cells in certain types of tissue.

In our case, we're using a dataset consisting of H\&E and P63 stained breast tissue images with the presence of metaplastic cancer. With cancer being among the top leading causes of disease-related death since 2016~\cite{Harding2018Peer19992016,SiegelMph2023Cancer2023}, prevention by screening and early detection becomes paramount to combat this disease.

\begin{figure}[htp] \centering

    \includegraphics[width=0.80\textwidth]{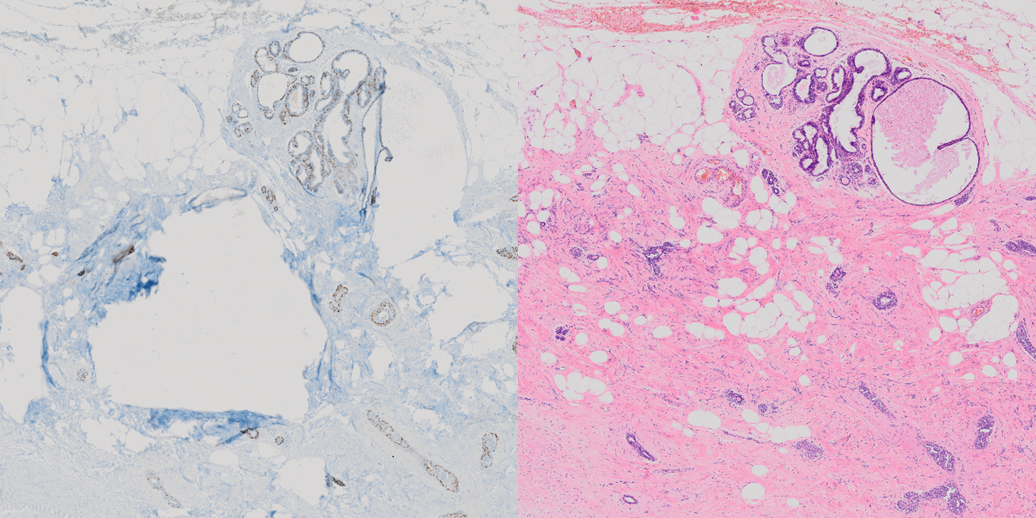}

    \caption{Demonstration of tissue damage in paired and aligned P63 stained tissue (left) compared with its H\&E counterpart (right). The issue is ripped in various places.}

    \label{fig:align}

\end{figure}

If an abnormal growth is found during inpatient screening for cancer, biopsies are a common next step in diagnosing the issue. This tissue is usually stained with Hematoxylin and Eosin (H\&E) to aid screening for cancerous growths. While H\&E stained tissue alone is sufficient and contains the necessary information to identify the presence of cancerous cells, this information is difficult to identify from the microscopy images. For this reason, immunohistochemical dyes are also used to help identify cancerous cells. Unfortunately, these dyes are often very expensive and also have the potential to damage the tissue, causing it to tear often, which is the case in our data as well. The effects of this are demonstrated in~\autoref{fig:align}.

For this reason, CycleGAN~\cite{Zhu2017UnpairedNetworks} and its derivatives are popular in this field for unpaired unsupervised training, and we used a modified CycleGAN architecture in our case as well called xAI-CycleGAN~\cite{sloboda2023xai}. We demonstrate its superior image quality in comparison to other existing methods.

\paragraph*{Approach}
We utilize xAI-CycleGAN with distilled features from the training data stored in the interpretable latent variable. We enhance this approach with an editable generation output on the basis of the interpretable latent variable using a modified semantic factorization algorithm, and introduce context loss to better preserve tissue structure and produce high quality images.

The output editing opens the doors to new possibilities in terms of the ability to control the generation of output using an easy-to-use tool that could serve to produce more data in cooperation with histopathologists.

We quantitatively evaluate our results by asking histopathologists to do a pair-wise comparison task in a survey to attempt to discern between a real and generated image, and then rate the realism of the image they believe not to be real. By this we show the approach has great potential but requires further work before it is usable in a medical setting. The code is available in~\href{https://github.com/slobodaapl/xAI-CycleGAN-2}{our GitHub repository}.

\paragraph*{Our Contribution}
\begin{itemize}
    \item xAI-CycleGAN enhanced with an editable output using the interpretable latent variable
    \item Introduction of context loss for improved preservation of tissue structure
    \item Improved output quality and reduced counterfactuals compared to previous solution
\end{itemize}

\section{Related work}
Various works have demonstrated relatively successful virtual staining of histopathological tissue~\cite{bai2023deep} and transformation from one type of stained tissue to another~\cite{bai2023deep,de2021deep,de2018stain}. Registration remains a consistent issue due to the nature of the problem, though it is possible. For paired transformation, various GAN architectures were used, from a simple GAN and DCGAN~\cite{Salimans2016ImprovedGANs}, as well as sequential cascaded GANs which show common use in the medical domain~\cite{CHEN2022107130}.

For the unpaired approach, CycleGAN~\cite{Zhu2017UnpairedNetworks} seems to be the dominant approach in nearly all cases. Therefore, we will look a bit closer at some related work that we predominantly use or relate to, below. 

\subsection{Overview of xAI-CycleGAN}
xAI-CycleGAN is an enhanced version of the CycleGAN~\cite{Zhu2017UnpairedNetworks} architecture that aims to improve the convergence rate and image quality in unsupervised image-to-image transformation tasks~\cite{sloboda2023xai}. It incorporates the concepts of explainability to provide a more powerful and versatile generative model.

One of the primary contributions of xAI-CycleGAN is the incorporation of explainability-driven training. Inspired by the work of Nagisetty et al.~\cite{Nagisetty2020XAI-GAN:Systems}, xAI-CycleGAN utilizes saliency maps from the discriminator to mask the gradients of the generator during backpropagation.

In addition, xAI-CycleGAN leverages the insights from the work of Wang M.'s Mask CycleGAN~\cite{Wang2022MaskVariable}, which introduces an interpretable latent variable using hard masks on the input. By combining these approaches, xAI-CycleGAN achieves enhanced explainability and convergence by taking advantage of information leakage into the interpretable latent variable.

One issue with this architecture is the production of counterfactuals by the generator and various artifacts and repeating patterns which may be worsened in the histopathological domain, and must be addressed.

We do this by introducing a new loss that attempts to preserve the structure and enforce context, by separating the context and style and forcing all style transformation to happen at the latent level of the generative model and preventing the encoder or decoder portion from doing any color transformation.

\subsection{SeFa algorithm for editable outputs}
The Closed-Form Factorization of Latent Semantics in GANs (SeFa) algorithm enabled the means to control the generated output in semantically interpretable ways~\cite{Zhou2021Closed-FormGANs}, such as the ability to modify the expression, color, posture, or other human-interpretable aspects.

The approach identifies semantically significant vectors in latent space with high variability, which are added to the activations from these layers by multiplying the eigenvector by an arbitrary factor.

Unfortunately, this approach only works with simpler generative networks that draw from noise distributions to generate images. In CycleGAN and other transformative generative models, the auto-encoder-like structure that starts and ends with an image does not have a 2-dimensional latent weight matrix, but instead a 4-dimensional one, making it impossible to apply this algorithm in its pure form.

We modify this approach to allow modification using the interpretable latent variable used in xAI-CycleGAN with some degree of success, but lose semantically interpretable aspects of this method since we only utilize one layer for this purpose due to difficulty of training and stability reasons.

Other methods also exist for image modification, such as the use of sequential GANs for editing~\cite{cheng2020sequential,pajouheshgar2022optimizing,collins2020editing} or inpainting, however all rely on a similar method as SeFa, with the difference of some using attention-based mechanisms~\cite{vaswani2017attention} to achieve this. The common problem still remains that these methods cannot be easily adapted to an auto-encoder-like architecture for unpaired image transformation, such as CycleGAN.

\subsection{cCGAN for stain transformation}
cCGAN (conditional CycleGAN)~\cite{xu2022ganbased} is a notable approach in the domain of stain transformation for histology images. It aims to translate images from one stain to another while preserving important structural information.

The cCGAN approach employs a conditional variant of the CycleGAN architecture, incorporating additional information such as staining information as input to guide the translation process.

While cCGAN has shown promise in transforming between stains, there are some limitations to consider. One challenge is the preservation of fine-grained structural details during the transformation process. Due to the absence of explicit constraints on structure preservation, cCGAN may struggle to accurately preserve important structural information, which is shown in their results.

We address the issue of structure preservation by incorporating context loss, which enforces consistency between encoded representations, and emphasizes the preservation of structural features during the transformation.

\section{Methods}
Here we will focus primarily on the context preserving approach by using context loss to improve carry-over of structural information and also fixing the artifacts and counterfactuals produced by xAI-CycleGAN. We also introduce a modified algorithm to edit the outputs of the generator to be able to adjust generation results to better match ground truth images of the matching domain.

\subsection{Dataset}
The dataset consists of 32 pairs of H\&E and P63 stained tissue samples of varying resolutions, all around 3-5 gigapixels in total size. These were cut into 1024 by 1024 pixel slices, and then downscaled to 256 by 256 which was the final resolution used during training.

Each slice was converted to LAB color space before being passed to the model, and checked by a combined method of luminosity-based histogram and image entropy to determine whether the image contained a sufficient amount of tissue, or whether it only consisted of background. This resulted in a total of 110'564 training samples used to train the model.

The tissue samples were obtained internally, and thus are not available publicly.

\subsection{Separating structure from style}
In order to preserve the structure of tissue and enforce only the style alone to change, we introduce the context-preserving loss, or just context loss. Context loss takes the encoded features of the same original image from the encoder of both generators, in both domains, and compares them to each other using Huber loss, and also compares the encoded features of the fake/output image on input of both generators across both domains.

This ensures that the encoder focuses entirely on the structural aspects of the image rather than style aspects, in order to ensure that no style conversion is done by the encoder and that the structure is identical between both the transformed image and the original image after the encoding step or also before the decoding step of the cycled images. This confirms that structure is preserved while only style changes, with most structure information being stored in the weights of the encoder and decoder step.

\begin{figure}[ht] \centering

    \frame{\includegraphics[width=0.5\textwidth]{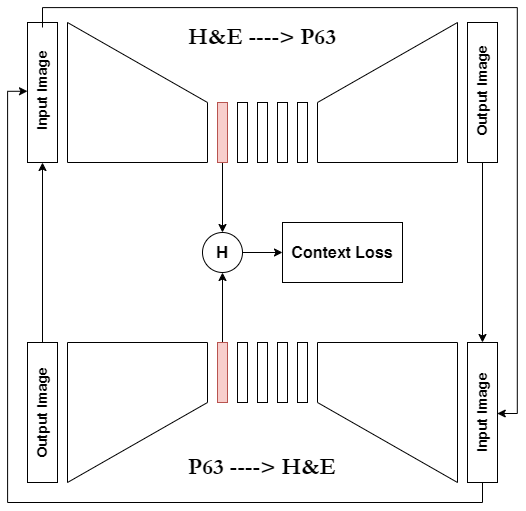}}

    \caption{Demonstration of the context loss computation. The encoded original image representation from both domains is compared to produce a partial loss, and added with the encoded fake/output representation of both domains in both generators to produce the final loss.}

    \label{fig:contextloss}

\end{figure}

\noindent
If we consider the H\&E to P63 generator encoder as a function $X_e()$ and the P63 to H\&E generator encoder as a function $Y_e()$, and then an original H\&E image to be $A$ and the generated counterpart $A'$, then a P63 original image to be $B$ and the generated counterpart $B'$, and finally the Huber loss function as $H()$, then we can define the context loss as follows:

\begin{equation}
    H(X_e(\alpha), Y_e(\beta)) = H_\gamma(\alpha, \beta)
\end{equation}

\begin{equation} \label{eq:context}
    \mathcal{L}_{context} = \frac{H_\gamma(A, A) + H_\gamma(B, B)}{2} + \frac{H_\gamma(A', B) + H_\gamma(B', A)}{2}
\end{equation}

\vspace{0.25cm}
\noindent
This loss is further adjusted by a parameter to weigh its importance in the total objective function for the generators, which is a hyperparameter. Our context loss enforces that the encoders should behave the same in both generators regardless of the input image, in order to preserve the structure. This technically means that after training only a single encoder and decoder is needed for both domains with only the latent transformation step being separate.

\subsection{Editable generation results using SeFa}
We utilize the SeFa algorithm to extract important eigenvectors from the interpretable latent variable in xAI-CycleGAN, akin to the method in~\cite{Zhou2021Closed-FormGANs} and incorporate them into the generation process. The goal is to enable fine-grained editing of the generated images by manipulating specific attributes or semantics through the modification of the interpretable latent variable.

To apply SeFa in xAI-CycleGAN, we first extract the most significant eigenvectors from the interpretable latent variable. Once the important eigenvectors are obtained, we incorporate them back into the weights of the corresponding layer during the forward pass of the network. This is achieved by multiplying the most significant eigenvector with its transposed self to produce a matrix of the same size as the weight matrix of the interpretable latent variable.

\begin{equation} \label{eq:wsum}
    W^* = W+\sum_{i=j}^{k}{\eta_i^T \eta_i \cdot m_i}
\end{equation}

\noindent
In~\autoref{eq:wsum}, $W^*$ represents the resulting weight matrix for the forward pass, with $W$ being the base one, then $\eta_i$ represents the i-th eigenvector of $W$, and $m_i$ is the m-th multiplicative factor for the given eigenvector. 

This is the only modification to the algorithm and presents a unique approach which allows us to use only the single layer at a cost of losing the semantically interpretable directions, but allowing us to utilize this algorithm in CycleGAN, an entirely new context for the algorithm, only made possible by the intentional leakage of semantically important information into the interpretable latent variable.

We allow a choice of any range of top 16 most important eigenvectors, where all are applied to the matrix with the same multiplicative factor in order.

\begin{figure}[ht] \centering

    \frame{\includegraphics[width=0.60\textwidth]{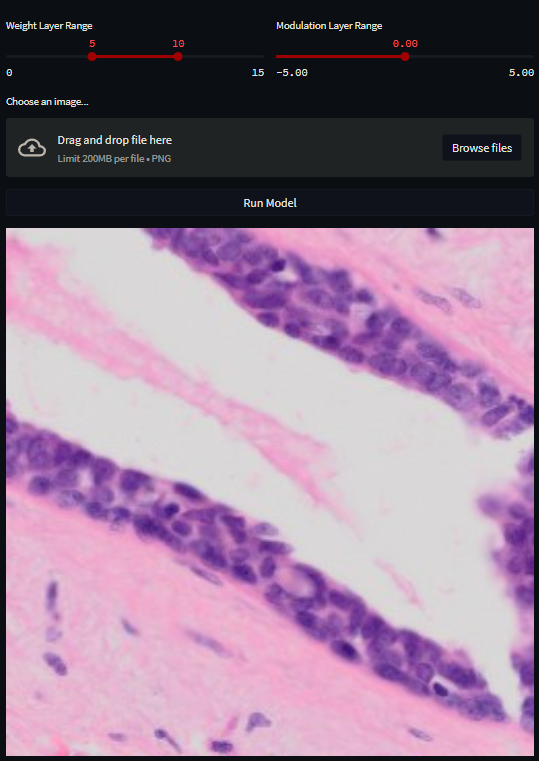}}

    \caption{The user interface to manipulate the image in real time before the image is transformed. Currently, a base H\&E image is shown, not yet transformed. For simplicity, we use a single multiplicative factor for every range of eigenvectors due to only using one layer, as such there would be little benefit to manually manipulate each factor, due to lack of semantically interpretable directions with the lack of interpretable latent variable layers.}

    \label{fig:strl1}

\end{figure}

We developed an easy-to-use tool using Streamlit shown in~\autoref{fig:strl1} to allow modifying generated outputs on the fly and explore the effects of this output editing, and to allow to interactively edit the images to more closely resemble the ground truth images.

\section{Results}
Using the already well-performing xAI-CycleGAN further enhanced with context loss, we vastly increase the visual fidelity and level of structure preservation in the images, while also preventing any counterfactuals or repeating patterns from appearing, which was an issue shown in xAI-CycleGAN.

\begin{figure}[htp] \centering

    \frame{\includegraphics[width=0.75\textwidth]{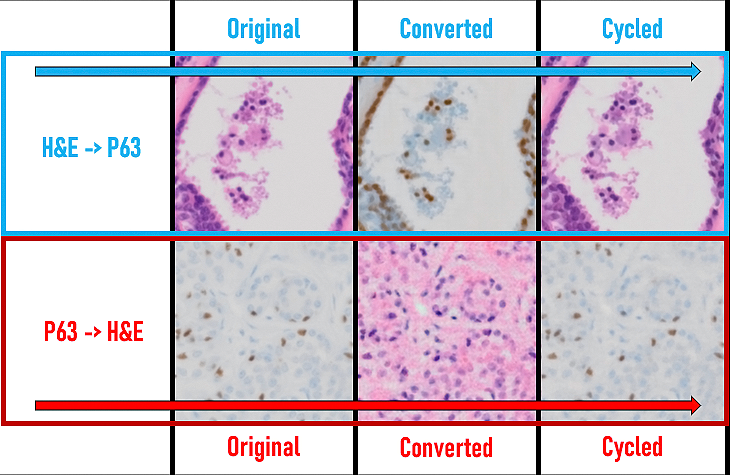}}

    \caption{Example of results generated from the model in both transformation directions, showing very high quality of outputs with preserved tissue structure.}

    \label{fig:res}

\end{figure}

We observe consistently high quality outputs for a variety of inputs seen in~\autoref{fig:res}, however upon closer examination with the help of a histopathologist, there are inconsistencies or impossible arrangements in the produced tissue based on the coloring of the cells with the P63 dye, which shows the model was not able to fully and correctly learn the complexities of how the tissue should look, but it has shown relatively good attempts to correctly identify myoepithelial cells which are colored brown in the P63 stained images.

\begin{figure}[htp] \centering

    \frame{\includegraphics[width=0.75\textwidth]{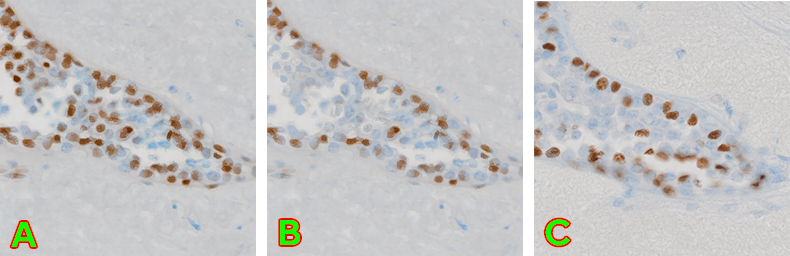}}

    \caption{A demonstration of successful editing capabilities of the tool. \textbf{A} contains image converted from H\&E to P63 using the tool, without any modifications applied. \textbf{B} contains an image with modifications applied using the tool, to best match the real image. \textbf{C} contains the unmodified original P63 image of the same region.}

    \label{fig:showcase}

\end{figure}

In some cases, we were able to use our new editing approach to modify the outputs of the generator in order to replicate the ground truth appearance relatively well, for instance in~\autoref{fig:showcase}.

\noindent

As visible in~\autoref{fig:showcase}, the method can be successfully applied to get relatively close to the real counterpart. It is not perfect due to the differences between the two samples used to prepare the H\&E stained tissue and P63 stained tissue.

In addition to qualitative assessment, we also asked 4 histopathologists to take a survey with a pair-wise comparison task to identify the real image from two images, the other image being generated and edited using our method to best match the real counterpart. The histopathologist was also asked to rate the image they believed was generated/fake on a scale from 1 (not realistic) to 6 (very realistic).

In the case where an image was identified incorrectly, we considered the score to be 6 in order to quantitatively assess the performance of the model. These results are demonstrated in~\autoref{tab:survey}.

\begin{table}[H]
\caption{A total of 8 image pairs were presented to 4 histopathologists; A generated image and a ground truth matched image. Each histopathologist was asked to guess the real image, and rate how realistic the generated image looks from 1 to 6. Incorrect guesses were assigned a score of 6 for realism of the generated image. 65.62\% of the images were determined correctly, which is relatively close to the ideal 50\% which would mean a complete inability to tell generated and real images apart.}
\label{tab:survey}
\frame{
\centering
\resizebox{\textwidth}{!}{%
\begin{tabular}{c|c|c|c}

\multicolumn{1}{l|}{} & Correctly Identified & Incorrectly Identified & Average Realism Rating \\ \hline
Pair 1        & 2  & 2  & 5.5  \\ \hline
Pair 2        & 2  & 2  & 5.25 \\ \hline
Pair 3        & 2  & 2  & 5.25 \\ \hline
Pair 4        & 3  & 1  & 5    \\ \hline
Pair 5        & 3  & 1  & 4.75 \\ \hline
Pair 6        & 3  & 1  & 5    \\ \hline
Pair 7        & 2  & 2  & 4.75 \\ \hline
Pair 8        & 4  & 0  & 4    \\ \specialrule{.2em}{.1em}{.1em} 
\textbf{Total/Average} & \textbf{21} & \textbf{11} & \textbf{4.94} (max 6)
\end{tabular}}
}
\end{table}

\noindent
We presented a total of 8 image pairs to 4 histopathologists. The results show a \textbf{total averaged realism score of 4.94} which is well above the midpoint, where the histopathologists were able to \textbf{correctly identify the real image 65.62\% of the time}. The histopathologists commented that the visual quality is very high, though without a broader investigation of the whole converted tissue sample it is impossible to truly judge the performance of the model, thus more research is needed, and we see potential for further improving the method.

\section{Discussion}
The results obtained from xAI-CycleGAN demonstrate its potential for advancing the field of stain transformation. The structure preservation and editable outputs offer exciting prospects for improving the accuracy and realism of unsupervised image-to-image transformation tasks. However, despite the promising results, there is still considerable room for improvement before xAI-CycleGAN can be effectively deployed in a medical setting.

One area of further research lies in optimizing the editing capabilities of xAI-CycleGAN. While the current approach allows for fine-grained editing by manipulating the interpretable latent variable, there is still a need to identify the optimal editing settings for different images. A potential avenue is training a mini-network on top of xAI-CycleGAN using the data produced from the editing approach. This mini-network could learn to identify the correct editing settings based on the input image, leading to consistently good results. By automating the selection of editing parameters, the system becomes more user-friendly and efficient, reducing the reliance on manual adjustments.

Moreover, expanding the editing approach to include more interpretable layers could provide better control over semantically meaningful transformations. Currently, the editing process primarily focuses on modifying the interpretable latent variable, which may lack direct semantic interpretability. By incorporating additional interpretable layers into the editing framework, it becomes possible to manipulate specific attributes or semantics directly, leading to more intuitive and meaningful editing capabilities. This expansion would offer greater control and customization in generating histology images that closely resemble the ground truth.

While xAI-CycleGAN shows promising potential, several considerations need to be addressed before its practical adoption in medical settings. Robust evaluation on larger and diverse datasets is necessary to validate its performance across various staining protocols and histology image types. The integration of expert knowledge and feedback from histopathologists is crucial for refining the model and ensuring that the transformed images maintain diagnostic relevance and accuracy. Additionally, efforts to further optimize and fine-tune the architecture, loss functions, and training procedures are essential to maximize the model's performance and generalizability.

With further research and development, xAI-CycleGAN holds promise for enhancing the accuracy, efficiency, and clinical utility of histology image analysis in medical practice, as evidenced by the positive feedback from histopathologists that participated in our survey.

\section{Future work}
Our present endeavor has laid the foundation for advancements in the field of virtual staining in histopathology, but there are aspects we seek to enhance in future work.

We plan to evaluate our xAI-CycleGAN architecture using a separate, diverse dataset. The intent is not only to affirm the robustness and adaptability of our model, but also to identify specific areas of potential improvements. An additional dataset will provide a more extensive ground for testing the limits and capabilities of our approach. With the extra information, we could better assess our approach's generality and reliability, and adjust our algorithm accordingly.

Additionally, as our model possesses the unique capability of editable outputs, we will explore how we can optimize this feature to improve the generated images. We aim to implement an iterative feedback loop where the edits are used to fine-tune the model, subsequently enhancing the generated outputs.

By moving ahead in these directions, we anticipate refining our approach to create a more precise and widely applicable tool for histopathological stain transformations. The ultimate goal is to establish a system that offers superior performance in generating virtual stains, thus offering the possibility of substantial savings in time, resources, and cost while maintaining diagnostic accuracy.

\newpage
\bibliographystyle{splncs04}
\bibliography{references}

\begin{thebibliography}{10}
\providecommand{\url}[1]{\texttt{#1}}
\providecommand{\urlprefix}{URL }
\providecommand{\doi}[1]{https://doi.org/#1}

\bibitem{bai2023deep}
Bai, B., Yang, X., Li, Y., Zhang, Y., Pillar, N., Ozcan, A.: Deep
  learning-enabled virtual histological staining of biological samples. Light:
  Science \& Applications  \textbf{12}(1), ~57 (2023)

\bibitem{de2018stain}
de~Bel, T., Hermsen, M., Kers, J., van~der Laak, J., Litjens, G.:
  Stain-transforming cycle-consistent generative adversarial networks for
  improved segmentation of renal histopathology  (2018)

\bibitem{CHEN2022107130}
Chen, H., Yan, S., Xie, M., Huang, J.: Application of cascaded gan based on ct
  scan in the diagnosis of aortic dissection. Computer Methods and Programs in
  Biomedicine  \textbf{226},  107130 (2022).
  \doi{https://doi.org/10.1016/j.cmpb.2022.107130},
  \url{https://www.sciencedirect.com/science/article/pii/S0169260722005119}

\bibitem{cheng2020sequential}
Cheng, Y., Gan, Z., Li, Y., Liu, J., Gao, J.: Sequential attention gan for
  interactive image editing. In: Proceedings of the 28th ACM international
  conference on multimedia. pp. 4383--4391 (2020)

\bibitem{collins2020editing}
Collins, E., Bala, R., Price, B., Susstrunk, S.: Editing in style: Uncovering
  the local semantics of gans. In: Proceedings of the IEEE/CVF Conference on
  Computer Vision and Pattern Recognition. pp. 5771--5780 (2020)

\bibitem{de2021deep}
de~Haan, K., Zhang, Y., Zuckerman, J.E., Liu, T., Sisk, A.E., Diaz, M.F., Jen,
  K.Y., Nobori, A., Liou, S., Zhang, S., et~al.: Deep learning-based
  transformation of h\&e stained tissues into special stains. Nature
  communications  \textbf{12}(1), ~4884 (2021)

\bibitem{Harding2018Peer19992016}
Harding, M.C., Sloan, C.D., Merrill, R.M., Harding, T.M., Thacker, B.J.,
  Thacker, E.L.: {Peer Reviewed: Transitions From Heart Disease to Cancer as
  the Leading Cause of Death in US States, 1999–2016}. Preventing Chronic
  Disease  \textbf{15}(12) (2018). \doi{10.5888/PCD15.180151},
  \url{/pmc/articles/PMC6307835/ /pmc/articles/PMC6307835/?report=abstract
  https://www.ncbi.nlm.nih.gov/pmc/articles/PMC6307835/}

\bibitem{Nagisetty2020XAI-GAN:Systems}
Nagisetty, V., Graves, L., Scott, J., Ganesh, V.: {xAI-GAN: Enhancing
  Generative Adversarial Networks via Explainable AI Systems} (2 2020).
  \doi{10.48550/arxiv.2002.10438}, \url{https://arxiv.org/abs/2002.10438v3}

\bibitem{pajouheshgar2022optimizing}
Pajouheshgar, E., Zhang, T., S{\"u}sstrunk, S.: Optimizing latent space
  directions for gan-based local image editing. In: ICASSP 2022-2022 IEEE
  International Conference on Acoustics, Speech and Signal Processing (ICASSP).
  pp. 1740--1744. IEEE (2022)

\bibitem{Salimans2016ImprovedGANs}
Salimans, T., Goodfellow, I., Zaremba, W., Cheung, V., Radford, A., Chen, X.:
  {Improved Techniques for Training GANs}. In: Lee, D., Sugiyama, M., Luxburg,
  U., Guyon, I., Garnett, R. (eds.) Advances in Neural Information Processing
  Systems. pp. 2234--2242. Curran Associates, Inc. (2016)

\bibitem{Zhou2021Closed-FormGANs}
Shen, Y., Zhou, B.: {Closed-Form Factorization of Latent Semantics in GANs}.
  In: Proceedings of the IEEE/CVF Conference on Computer Vision and Pattern
  Recognition (CVPR). pp. 1532--1540 (2021)

\bibitem{SiegelMph2023Cancer2023}
Siegel~Mph, R.L., Miller, K.D., Sandeep, N., Mbbs, W., Ahmedin, ., Dvm, J.,
  Siegel, R.L.: {Cancer statistics, 2023}. CA: A Cancer Journal for Clinicians
  \textbf{73}(1),  17--48 (1 2023). \doi{10.3322/CAAC.21763},
  \url{https://onlinelibrary.wiley.com/doi/full/10.3322/caac.21763
  https://onlinelibrary.wiley.com/doi/abs/10.3322/caac.21763
  https://acsjournals.onlinelibrary.wiley.com/doi/10.3322/caac.21763}

\bibitem{sloboda2023xai}
Sloboda, T., Hudec, L., Bene{\v{s}}ov{\'a}, W.: xai-cyclegan, a
  cycle-consistent generative assistive network. arXiv preprint
  arXiv:2306.15760  (2023)

\bibitem{vaswani2017attention}
Vaswani, A., Shazeer, N., Parmar, N., Uszkoreit, J., Jones, L., Gomez, A.N.,
  Kaiser, {\L}., Polosukhin, I.: Attention is all you need. Advances in neural
  information processing systems  \textbf{30} (2017)

\bibitem{Wang2022MaskVariable}
Wang, M.: {Mask CycleGAN: Unpaired Multi-modal Domain Translation with
  Interpretable Latent Variable} (5 2022). \doi{10.48550/arxiv.2205.06969},
  \url{https://arxiv.org/abs/2205.06969v1}

\bibitem{xu2022ganbased}
Xu, Z., Huang, X., Moro, C.F., Bozóky, B., Zhang, Q.: Gan-based virtual
  re-staining: A promising solution for whole slide image analysis (2022)

\bibitem{Zhu2017UnpairedNetworks}
Zhu, J.Y., Park, T., Isola, P., Efros, A.A., Research, B.A.: {Unpaired
  Image-To-Image Translation Using Cycle-Consistent Adversarial Networks}
  (2017), \url{https://github.com/junyanz/CycleGAN.}

\end{thebibliography}

\end{document}